%% file: template.tex
\documentclass{vgtc} 
\let\ifpdf\relax

\newcommand{\rpm}{\sbox0{$1$}\sbox2{$\scriptstyle\pm$}
  \raise\dimexpr(\ht0-\ht2)/2\relax\box2 }

\usepackage{amsmath,amssymb,amsthm,enumitem}
\usepackage{booktabs}
\usepackage{tabularx}
\usepackage{float}

\usepackage{mathptmx}
\usepackage{graphicx}
\usepackage{times}

\onlineid{230}

\vgtccategory{Research}

\vgtcinsertpkg

\title{Exploring Non-Reversing Magic Mirrors for Screen-Based Augmented Reality Systems}

\author{Felix Bork\thanks{e-mail:felix.bork@tum.de}\\ %
	\parbox{1.6in}{\scriptsize \centering Technische Universit\"at M\"unchen \\ Munich, Germany}
	\and Roghayeh Barmaki\thanks{e-mail:rl@jhu.edu}\\ %
	\parbox{1.6in}{\scriptsize \centering Johns Hopkins University, Baltimore MD, United States} %
	\and Ulrich Eck\thanks{e-mail:eck@tum.de}\\ %
	\parbox{1.6in}{\scriptsize \centering Technische Universit\"at M\"unchen \\ Munich Germany} %
	\vspace{0.15in}
	\and Pascal Fallavollita\thanks{e-mail:pfallavo@uottawa.ca}\\ %
	\parbox{1.6in}{\scriptsize \centering University of Ottawa \\ Ottawa Canada} %
	\and Bernhard Fuerst\thanks{e-mail:pfallavo@uottawa.ca}\\ %
	\parbox{1.6in}{\scriptsize \centering Johns Hopkins University, Baltimore MD, United States} %
	\and Nassir Navab\thanks{e-mail:nassir.navab@tum.de}\\ %
	\parbox{3in}{\scriptsize \centering Technische Universit\"at M\"unchen, Munich Germany \\ Johns Hopkins University, Baltimore MD, United States}}

\input{0_abstract}

\CCScatlist{ 
\CCScat{H.5.1}{Information Interfaces and Presentation}
{Multimedia Information Systems}%
{Artificial, augmented, and virtual realities};
\CCScat{H.5.2}{Information Interfaces and Presentation}%
{User Interfaces}
{Ergonomics};
}

\begin{document}


\firstsection{Introduction}
\maketitle

\input{1_introduction}
\input{2_related_work}
\input{3_user_study}
\input{4_results}
\input{5_discussion}
\input{6_conclusion_future_work}

\acknowledgements{
We thank Anna-Maria von der Heide, Severine Habert, and Alexander Keppler for their help during the user study design and execution. Anatomy models were obtained from http://www.plasticboy.co.uk.}

\bibliographystyle{abbrv}
\bibliography{template}
\end{document}

%% file: 0_abstract.tex
\abstract{
Screen-based Augmented Reality (AR) systems can be built as a window into the real world as often done in mobile AR applications or using the Magic Mirror metaphor, where users can see themselves with augmented graphics on a large display. Such Magic Mirror systems have been used in digital clothing environments to create virtual dressing rooms, to teach human anatomy, and for collaborative design tasks. The term Magic Mirror implies that the display shows the users’ enantiomorph, i.e. the mirror image, such that the system mimics a real-world physical mirror. However, the question arises whether one should design a traditional mirror, or instead display the “true mirror” image by means of a non-reversing mirror? This is an intriguing perceptual question, as the image one observes in a mirror is not a ‘real view’, as it would be seen by an external observer, but a reflection, i.e. a ‘front-to-back’ reversed image. In this paper, we discuss the perceptual differences between these two mirror visualization concepts and present a first comparative study in the context of Magic Mirror anatomy teaching. We investigate the ability of users to identify the correct placement of virtual anatomical structures in our screen-based AR system for two conditions: a regular mirror and a non-reversing mirror setup. The results of our study indicate that the latter is more suitable for applications where previously acquired domain-specific knowledge plays an important role. The lessons learned open up new research directions in the fields of user interfaces and interaction in non-reversing mirror environments and could impact the implementation of general screen-based AR systems in other domains.}

%% file: 1_introduction.tex
``Mirror, mirror on the wall...". Throughout history, people have always been fascinated by surfaces in which they could see a reflection of themselves.
From the first ancient obsidian fragments, polished metals, and wealth-reflecting status objects of kings and aristocrats to the nowadays ubiquitous everyday item - mirrors have undergone a remarkable evolution accompanied by a profound impact on religion, sciences, and arts \cite{pendergrast2009mirror, melchior2002mirror}.
Many legends have sprung around mirrors and they are still subject to countless myths, superstitions, and rituals around the world.
According to one of those, mirrors are ``portals to another world or dimension".

\vspace{0.2cm}

\noindent On first thought, this might sound like a quote form a science-fiction story.
However, Bertamini argues that mirrors are in fact a window into a completely virtual world \cite{bertamini2010mirrors}. 
According to him, everything we see inside a mirror is completely virtual, which in a sense makes a mirror the perfect Virtual Reality (VR) system.
Our brain is tricked into thinking that people or objects we see inside a mirror physically exist.
Due to this phenomenon, mirrors have been the subject of extensive research in the fields of psychology and human perception. 
Especially the well known 'left-right-reversal' is a characteristic which immediately comes to the mind of most people when talking about mirrors.
Lifting your right hand in front of a mirror results in your enantiomorph lifting the left hand.
Contrary to this, a non-reversing mirror presents a view as seen by an external observer standing in front of the person.
Lifting your right hand now corresponds to your true mirror image also lifting the right hand.

\vspace{0.2cm}

\noindent By means of a video camera and a display device, it is possible to create a virtual mirror mimicking the properties of a physical mirror.
In the field of Augmented Reality (AR), such systems are referred to as Magic Mirrors as soon as virtual graphics are superimposed on top of the video image.
Magic Mirrors have been successfully employed in various disciplines, such as collaborative design tasks \cite{saakes2016mirror}, virtual shopping environments \cite{Wang2012}, and anatomy teaching \cite{Blum2012c}.
Up to now, all these Magic Mirrors have naturally adopted the mirror metaphor and designed the system to mimic a traditional, physical mirror.
To the best of our knowledge, neither has a non-reversing Magic Mirror system been published before, nor has work discussing the perceptual aspects of non-reversing Magic Mirrors.

\vspace{0.2cm}

\noindent In this paper, we fill this gap by investigating the perceptual differences between Magic Mirror and non-reversing Magic Mirror systems.
One prime application area in which a non-reversing Magic Mirror could be superior to a traditional Magic Mirror is human anatomy teaching.
By learning anatomy from textbooks and patient examinations, medical students and medical professionals are primed to having a third-person view onto the anatomy of someone else. 
However, they are not used to facing themselves in a mirror.
We designed a preliminary user study consisting of two experiments to investigate the ability of medical students to identify the correct placement of virtual, anatomical structures in both a traditional Magic Mirror and a non-reversing Magic Mirror setup.
The results presented in this paper indicate that in the case of anatomy teaching, a non-reversing Magic Mirror yields significant benefits over a traditional Magic Mirror.
Our results will help designing Magic Mirror systems in other domains, in which previous, domain-specific knowledge can be assumed.

\vspace{0.2cm}

\noindent In section \ref{sec:related}, we discuss related work on mirror perception and previously proposed Magic Mirror AR systems. Section \ref{sec:study} is dedicated to our user study and provides an overview of our employed research protocol and investigated conditions.
Subsequently, we present the results of the study in section \ref{sec:results} and discuss their relevance and implications in the context of general non-reversing Magic Mirror systems.

%% file: 2_related_work.tex
\section{Background \& Related Work} \label{sec:related} 
In order to explore the potential of a non-reversing Magic Mirror setup for AR applications, we first need to understand the perceptual phenomena occurring in real-life mirror environments.
We reviewed the human perception and psychology literature and discuss how these concepts can be transferred to an AR Magic Mirror context.
Furthermore, we analyzed previously proposed Magic Mirror systems and examine whether instead using a non-reversing mirror setup would be beneficial with respect to their intended purpose.

\subsection{General Mirror Perception}
Illusion of explanatory depth is a common phenomenon occurring in many different areas of our daily life \cite{rozenblit2002misunderstood}. People have the tendency to overestimate their knowledge of certain procedures or objects, as demonstrated by Lawson et al. on the example of bicycles \cite{lawson2006science}.
The same illusion of explanatory depth principle holds true for mirrors as well.
Even though we gaze into them multiple times a day, most of us do not have an in-depth understanding of what exactly happens on the surface of a mirror.
Betamini and Parks have shown during experiments, that people failed to accurately judge the size of their enantiomorph's face, especially with varying distance from the mirror \cite{Bertamini2005}.
In another study, Lawson and Bertamini provide evidence that people generally expect to see their own mirror image earlier when approaching a covered mirror from the side \cite{Lawson2006}.
Closely related to this research is the so called Venus effect \cite{Bertamini2003, bertamini2010venus}.
In painting, such as \emph{The Robeky Venus} by Vel\'azquez, it occurs if both a mirror and an actor looking into the mirror are present, with the reflection in the mirror not corresponding to the view of the actor but the view of the observer of the painting.
Most people describe such a painting as if the actor is looking at her own reflection in the mirror.

\vspace{0.2cm}

\noindent All of these misperceptions manifest the simultaneous simplicity and complexity of mirrors.
Especially the questions about \emph{why mirrors reverse left and right, but not up and down?} has been a controversial topic for decades \cite{gregory1998mirrors, Takano1998, corballis2000much, Takano2015}.
In mathematical terms, mirrors reverse across the axis perpendicular to their surface, such that front and back are reversed, similar to a glove being turned inside out.
This corresponds to a change of coordinate systems from a right handed to a left handed one (or vice versa).
However, Ittelson et al. showed that the reversal happens across the axis of greatest perceived symmetry \cite{Ittelson1991}.
Due to the bilateral symmetry of the human body, this axis coincided with the left-right axis.
Therefore, people tend to believe that their mirror image is formed by a rotation around the vertical (up-down) axis, i.e. by walking around the mirror to become the virtual self \cite{Bertamini2003a, bertamini2010mirrors}.
However, the resulting image is not how a regular enantiomorph looks like, but resembles the image produced by a non-reversing mirror.
Non-reversing mirrors show the true mirror image and can be built physically by placing two mirrors perpendicular to each other to form two sides of an equilateral triangle \cite{watson1986non}, or as proposed more recently, by combining a multitude of tiny mirrors to form a curved surface, which directs incoming light rays back to the observers eye such that the image is horizontally flipped \cite{hicks2012wide}.

\vspace{0.2cm}

\noindent In AR applications, implementing a virtual non-reversing mirror is as easy as rearranging the columns of a digital camera image.
However, to the best of our knowledge, no published work has yet discussed the possibility of employing the non-reversing mirror paradigm for AR.
In the following section, we explore the usage of mirrors in practice by reviewing previously proposed Magic Mirror systems in both AR and VR and discussing their potential to apply the non-reversing mirror paradigm.

\subsection{Interactive Mirrors}
Magic Mirrors have been widely used in fashion apparel simulations and virtual clothing. Kim and Cheeyoung \cite{kim2015augmented} presented a fashion coordination prototype that combined user recognition and the augmentation of face styles, make-up, glasses and dress fitting simulations in a mirror-like image representations.
For cosmetics and grooming, the Smart Makeup Mirror system \cite{iwabuchi2009smart} was introduced by Iwabuchi and Siio to facilitate and support wearing make up in form of a virtual dressing table. 
Two other examples were published by Rahman  et al. in form of a prototype assisting in the selection of cosmetic products \cite{rahman2010augmented}, and by Chu et al. who presented an advanced Magic Mirror jewelry shopping tool \cite{chu2010countertop}.
Another application area for Magic Mirrors are intelligent fitting rooms \cite{zhang2008intelligent}, superimposing virtual garments onto the user, e.g. shirts \cite{hilsmann2009tracking} or virtual shoes \cite{eisert20083}.
Zhang et. al presented an architecture for physical retail fitting rooms \cite{zhang2008intelligent}. 
%
%
Among all reported cosmetic and clothing Magic Mirrors, only the Smart Makeup Mirror system \cite{iwabuchi2009smart} implemented both a traditional Magic Mirror and a non-reversing one as two distinct views of the user. They stated that professional makeup artists always recommend the to validate a person's appearance from the viewpoint of another person standing in front using a non-reversing mirror.
Some scholars used virtual reality technology for the purpose of medical rehabilitation. 
Obdrzalek et al. \cite{obdrvzalek2012real} presented a real-time pose detection and tracking system from vision-based 3D data tele-rehabilitation in a virtual mirror environments. 
Furthermore, Kallman et al. \cite{kallmann2015vr} proposed a direct motion demonstration system showing an overlapped virtual avatar on a large display screen during rehabilitation exercises. 
Since the avatar mirrors the poses of the user in real-time, this system could be considered as a Magic Mirror and it uses a similar approach to that of \emph{mirror therapy} rehabilitation. 
Yavuzer et al. investigated the impact of Mirror Therapy (using a physical mirror for some upper limb exercises) on upper-extremity motor recovery inpatients with subacute stroke \cite{yavuzer2008mirror}. 
Another interesting application area for Magic Mirror systems is human anatomy education and training.
Several works have been published before \cite{Blum2012c,bauer2015living, MA2016}. 
Learning human anatomy is critical to medical students, but the learning process itself can be challenging. Students spend countless hours participating in traditional cadaver labs and studying physical anatomy models and atlases of anatomy. Such traditional education practices have changed relatively little in the last few decades, and while these practices have advantages, they also have significant drawbacks. 

\vspace{0.2cm}

\noindent In this work, we chose anatomy learning as a sample application. Left-right confusion is a topic medical students have to deal with a lot. Infamous errors occurred when e.g. the wrong kidney was removed, or a cut was made on the wrong side. Therefore, it is critical for an AR system to follow medical conventions from textbooks or other educational resources. This made us choose anatomy education as application for investigating the perceptual differences between traditional Magic Mirror and non-reversing Magic Mirror systems.

%% file: 3_user_study.tex
\section{User Study} \label{sec:study}
To investigate the potential of a non-reversing Magic Mirror (NRM) system and the benefits such a design could provide over a traditional Magic Mirror (MM), we implemented both of these visualizations in an AR anatomy learning demo application, enabling the augmentation of 3-dimensional organ models on top of the user standing in front of the system.
We designed a preliminary user study to compare the performance of medical students in identifying the correct placement of virtual anatomical structures in these two setups.
Two different experiments were conducted: 
in the first one, the virtual organs were augmented on top of the body of a second person facing the participant, creating the familiar patient examination view.
In the second experiment, both the NRM and RM conditions were presented to the participants, while 
they could see their mirror image augmented with virtual organs on the display device.
Throughout our user study, we used five different organ models for augmentation: the \emph{liver}, \emph{gallbladder}, \emph{colon}, \emph{pancreas}, and \emph{stomach}.
All of these organs can clearly be associated to either the left or right side of the human body.
In the following sections, we give a detailed overview of our hardware setup and the two experiments conducted during our preliminary user study. 
Furthermore, we describe the results of a pilot study preceding our main user study.

\subsection{Hardware Setup} \label{hardware}
In accordance with all of the previously mentioned AR Magic Mirror systems, the hardware components of our anatomy teaching application are a video camera and a large, 60 inch display device.
For the former, we chose the Microsoft Kinect V2 sensor which combines both an RGB and a depth camera in a single housing.
The Kinect was mounted on top of the display device at a height of two meters facing downwards with an angle of 17 degrees.
We positioned participants 150 cm away from the display device during the entire time of the user study. For experiment 1, the second person was positioned on the other side of the display device at the same distance as the participant.
For the purpose of augmenting virtual organs on top of the user, we employed the Kinect skeleton tracking API.
During the experiments, participants were asked to decide whether these organs are displayed on the anatomically correct side of the body or not.
This decision process was controlled by means of two buttons on a Logitech R400 presenter.
A third button on the bottom was programmed to switch to the next condition during the experiments, such that the entire experiment procedure was controlled by the participant.


\subsection{Experiment 1 - See-Through Window}
In experiment 1, we used a see-through window (STW) approach instead of designing a RM or NRM setup. 
Both the participant and the second person volunteering in the user study were positioned on opposite sides of the display device, with the Kinect sensor facing the volunteer.
Using this system setup, participants were presented with an observatory view known to them from countless patient examinations and from anatomy textbooks.
During the course of the experiment, the aforementioned five organs were augmented onto the body of the volunteer, either on the anatomically correct side or not.
Therefore, two different conditions were subject to investigation during experiment one: 
\begin{enumerate}
  \item \textbf{STW-NF}: See-Through Window, Organs Not Flipped
  \item \textbf{STW-F}:~~~~See-Through Window, Organs Flipped
\end{enumerate}
The first condition (STW-NF) corresponded to the anatomically correct placement.
This experiment had 10 different conditions ($1$ (NRM view) $\times 2$ (organs not flipped vs. flipped views) $\times 5$ (organ types), and two trials or repetitions, such that each participant made a total of 20 decisions during the experiment 1.
We employed a balanced Latin square matrix for randomizing these conditions across the study participants \cite{williams1949experimental}.

\vspace{0.2cm}

\noindent The goals of experiment 1 are three-fold:
\emph{i)} making participants familiar with the entire system and decision procedure; \emph{ii)} verify that they have enough anatomical knowledge to successfully participate in the user study. People who did not manage to answer at least $80\%$ of the questions correct were excluded from the user study evaluation; and \emph{iii)} confirming that medical students are comfortable with the patient examination view from their studies and do not have any difficulties in providing correct answers for this setup.

\subsection{Experiment 2 - Non-Reversing vs. Reversing Mirror}
After successful completion of the eligibility test in experiment 1, participants moved over to the main part of the user study in form of experiment 2. 
We turned the Kinect sensor back to its original position to face the participant.
The same five virtual organs were now augmented onto the participant's body and he was asked to choose whether the placement is anatomically correct or not.
Four different conditions were traversed:
\begin{enumerate}
  \item \textbf{NRM-NF}: Non-Reversing Mirror, Organs Not Flipped
  \item \textbf{NRM-F}:~~~ Non-Reversing Mirror, Organs Flipped
  \item \textbf{RM-NF}:~~~ Reversing Mirror, Organs Not Flipped
  \item \textbf{RM-F}:~~~~~~ Reversing Mirror, Organs Flipped
\end{enumerate}

\noindent Similar to experiment 1, the conditions for which the organs were flipped (NRM-F \& RM-F) corresponded to an anatomically wrong placement.
Experiment 2 had a $2$ (NRM vs. RM views) $\times  2$ (organs not flipped vs. flipped views) $\times 5$ (organ types) within-subjects design with two trials, leading to $20$ different conditions and $40$ ($20$ conditions $\times 2$ trials) total decisions for each participant. Again, the experiment was fully randomized using a balanced Latin square matrix.
After the participants provided an answer by means of the hand held presenter, we displayed a black screen and the participant was asked to continue with the next condition by pressing a button on the presented. 
Only after this, the camera image became visible again.
This design choice was made in order to avoid too obvious switches between the NRM and RM conditions.

\vspace{0.2cm}

\begin{figure*}[h]
  \centering
  \includegraphics[width=2.1\columnwidth]{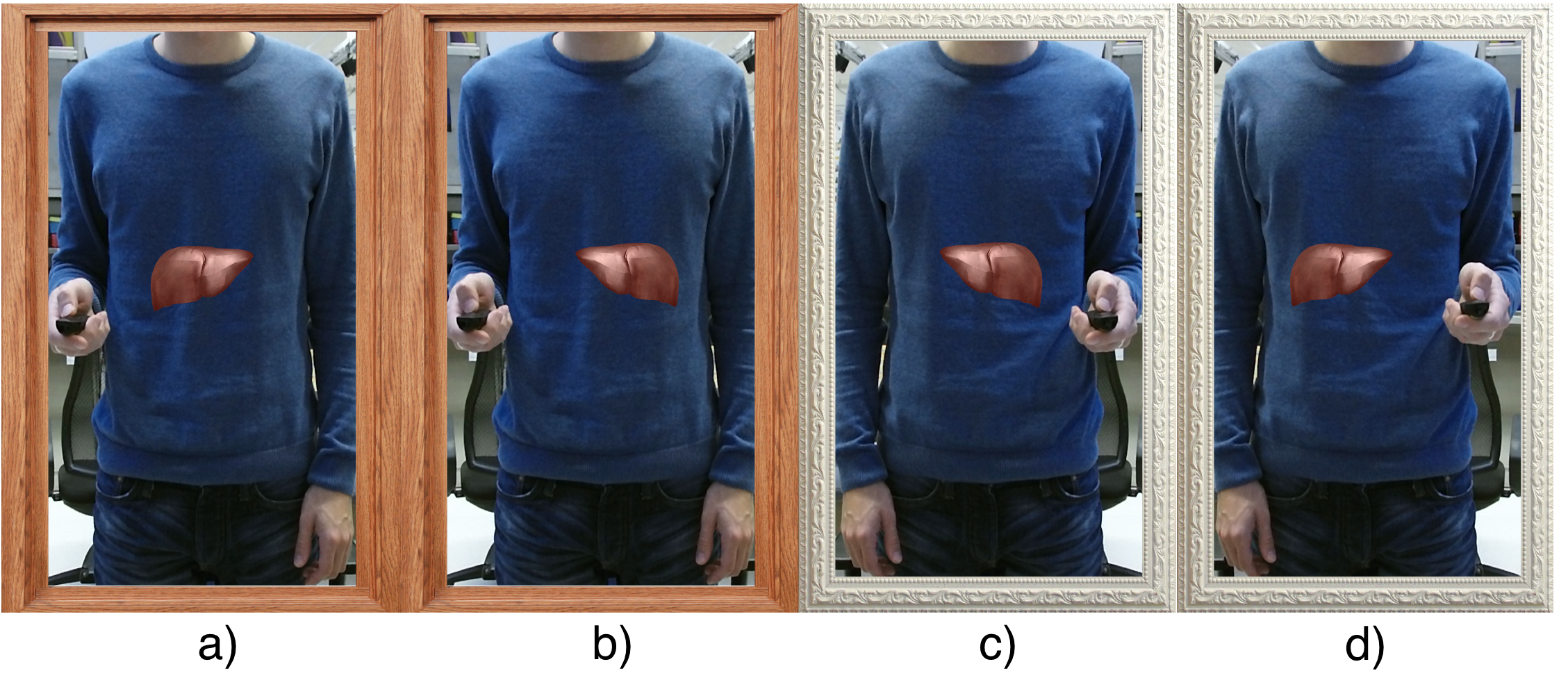}
  \caption{Comparison of the four different user study conditions: a) NRM-NF, b) NRM-F, c) RM-NF, d) RM-F. The participant is lifting the right hand and chooses by means of a pointer device whether the presented augmentation is anatomically correct. For the anatomically correct conditions a) and c), the liver is augmented on the right side of the body.}
  \label{fig:conditions}
\end{figure*}

\noindent The main goal of experiment 2 was to study whether an NRM provides perceptual benefits over the traditional RM design and whether these in return yield an increased overall rate of correct answers. An AR view of all four different conditions is shown in Fig. \ref{fig:conditions}.

\subsection{Pilot Study} \label{subsec:Pstudy}
We conducted a pilot study (N=4, 1 F) with advanced medical students from a surgical clinic of the affiliated university. Two of them were in their fourth medical year finishing their clinical rotations, while the other two were juniors.
Our goals for the pilot study were to verify the feasibility of our study protocol, identifying potential flaws and areas of confusion, and finally to come up with hypotheses for the subsequent full scale user study.

%

\vspace{0.2cm}

\noindent All participants achieved sufficiently high scores in experiment 1, such that none of them would have been excluded from the statistical analysis during our main user study.
%
%
Examining the percentage of correct answers showed a much better overall performance for the two NRM conditions.
The fourth-year senior participants completely failed to answer correctly for the RM-condition questions while the other two (juniors) performed comparably well for both NRM and RM visualizations. In a post-experiment interview, the two former participants mentioned that they solely focused on the location of virtual organs, therefore ignoring the difference between the NRM and RM conditions completely. Further elaboration about this observation is reported in the discussion section. 
Though the system should be intuitively usable without any introduction, we decided to improve upon the design of our study protocol by explicitly introducing the two different views (NRM and MM) to the participants prior to experiment 2 for the main user study.
That way, we can ensure that all participants understand the two visualizations and take the current view into consideration when deciding whether a virtual organ is placed correctly or not.

\vspace{0.2cm}

\noindent Based on the findings during our pilot study, we formulated the following four hypotheses subject to statistical tests in the subsequent full-scale user study:
\begin{enumerate}[leftmargin=*,labelindent=10pt,label=\bfseries H \arabic*.]
  \item The percentage of correctly identified virtual organs is higher for the NRM conditions in comparison to the regular RM conditions among the medical students with anatomical pre-knowledge.
  \item The average decision or response time (in seconds) is smaller for the NRM conditions compared to the RM conditions for medical students.
  \item All of the five augmented organs perform similar in terms of correct placement identification among all the participants and across the four main conditions of the study.
  \item The seniority level of participants does have an impact on the percentage of correctly identified virtual organs.
\end{enumerate}
Furthermore, we expect the vast majority of participants to prefer the NRM conditions and to quickly establish the link between these views and the familiar patient examination and textbook view.

\subsection{Main Study}
In order to test the hypotheses listed in the previous section, we recruited 21 medical students (10M, 11F), almost half of which (N=10) were in their fourth year, while the other half were distributed equally among the second and third year.
We excluded one male, second-year student from the study who did not manage to reach an average of at least $80$\% correct answers in experiment 1.
Therefore, the total number of participants was 20 (9M, 11F and 10 Senior, 10 Junior).
None of the pilot study participants took part in the main user study.

\vspace{0.2cm}

\noindent Participants were shown examples of NRM and RM conditions to familiarize themselves with the system. Furthermore, the previously discussed improvement to the user study protocol in the form of explicitly presenting the type of view for both the NRM and RM conditions was considered in this experiment.
Besides these slight changes, we kept the user study protocol and the hardware setup the same for experiments 1 and 2.

%% file: 4_results.tex
\section{Results} \label{sec:results}

In this section, we calculate the correlations of task completion times and correct answers with the independent variables of the user study and report the results. 
The average percentage of correct answers in experiment 1 was beyond 90\% (STW-NF: $94.5 \%$; STW-F: $92 \%$), confirming that all participants had sufficient knowledge about the human anatomy to verify the correct placement of all 5 organs of interest (liver, colon, gallbladder, pancreas, and stomach).
The average of decision times for both of the two conditions in experiment 1 were almost identical (STW-NF: $3.78 \rpm 2.06$ s; STW-F: $3.85 \rpm 2.07$ s), see table \ref{tab:evaluation_results}.
On average, participants had shorter decision times for the liver ($3.26$ s), stomach ($3.34$ s), and colon ($3.45$ s), compared to the gallbladder ($3.94$ s) and pancreas ($5.01$ s).
This was accompanied by a slightly higher average of correct answers for the former three organs (liver: $97.5\%$, stomach \& colon: $95\%$; pancreas: $90\%$; gallbladder: $88.75\%$).
\vspace{0.2cm}

\begin{figure*}[h]
  \centering
  \includegraphics[width=2.1\columnwidth]{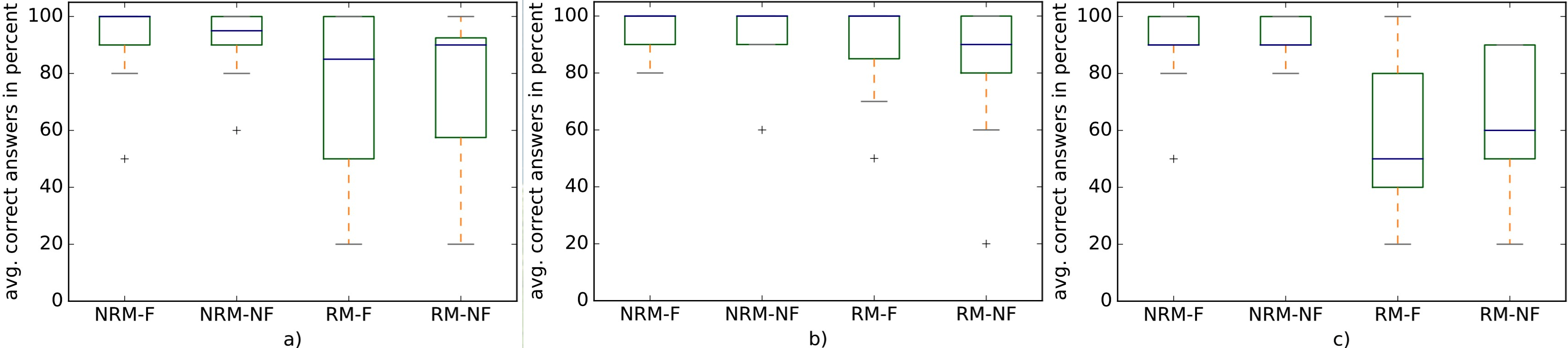}
  \caption{Comparison of average correct answers given for the task of identifying virtual organs superimposed on the participants bodies for the two NRM and two RM conditions for a) all participants, b) junior participants, and c) senior participants.}
  \label{fig:correct-answers}
\end{figure*}

\begin{figure}[h]
  \centering
  \includegraphics[width=1\columnwidth]{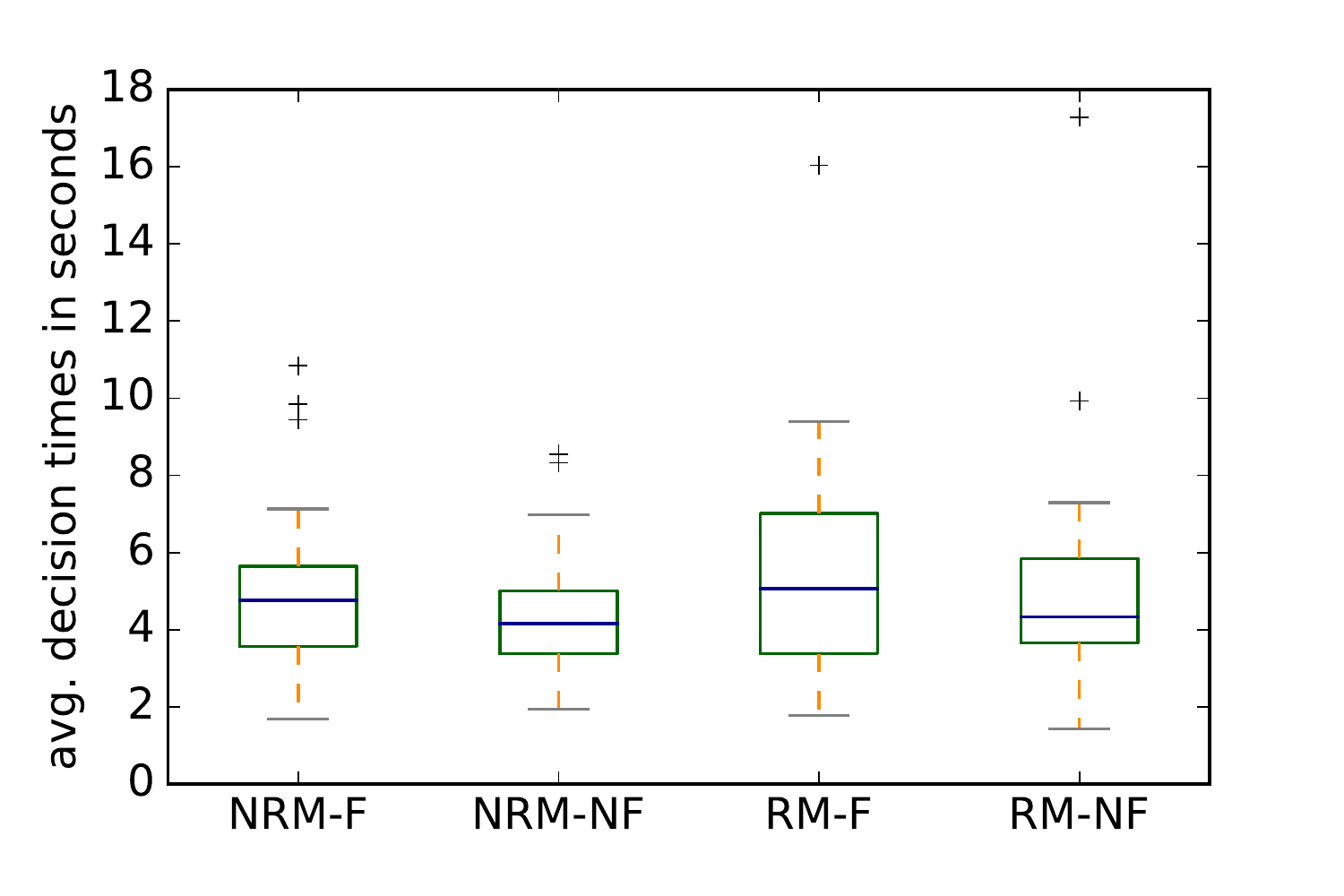}
  \caption{Comparison of average decision times from experiment 2 for the 4 underlying conditions.}
  \label{fig:completion-times}
\end{figure}

\noindent In order to investigate the differences between an NRM and a traditional RM design in anatomy learning application and to test our previously formulated hypotheses, we analyzed results obtained from experiment 2.
Figure \ref{fig:correct-answers} a) shows the average percentage of correct answers of all four conditions among participants. There were large deviations for the RM conditions, compared to very small deviations for the NRM conditions.
The average of correct answers for both NRM conditions was $92.25\%$ compared to only $75.5\%$ for the two RM conditions. This difference was statistically significant $(F_{1,19} = 10.92$, \emph{p} $< .005$), see also Fig. \ref{fig:correct-answers} a) by comparing the two left box-plots with the other two.
We chose a similar approach for analyzing the average decision times among all 20 participants for experiment 2, summarized in Fig. \ref{fig:completion-times}. 
It shows that participants were slightly faster in the two NRM conditions and were relatively slow for the RM-F condition. 
The mean decision times were $4.77$ s for both NRM conditions and $5.49$ s for the RM conditions. As there was substantial variation in the observations across participants, the difference was not statistically significant as revealed in an analysis of variances $(F_{1,19} = 1.56$, ns).  
An overview of the average of decision times for the 4 individual conditions can be obtained from table \ref{tab:evaluation_results}.
Comparing the percentage of correct answers for each of the 5 different organs among all participants revealed comparable numbers to those of experiment 1. In the NRM conditions, the pancreas was the best performing organ ($97.5$ \% correct answers), followed by the liver ($93.75$ \%), colon and gallbladder ($91.25$ \%), and the stomach ($88.75$ \%).
Overall results for the RM conditions were worse. Especially the stomach was identified as correctly placed only in $65$ \% of all cases.
The other percentages were: liver ($75$ \%), gallbladder \& pancreas ($78.75$ \%), and the colon ($80$ \%).

\vspace{0.2cm}

\noindent Lastly, we examined whether the seniority level of participants affects the outcome of the user study. 
We split all participants into two groups: 
To the first group, hereafter referred to as the \emph{junior group}, all medical students in their second or third year were assigned.
Consequently, all fourth year medical students formed the \emph{senior group}. The two groups were balanced and both contained 10 participants. For correct answer percentages, we observed an interesting difference among juniors and seniors. Juniors performed significantly better for the RM conditions than seniors $(F_{1,18} = 5.09$, \emph{p} $< .05$). Furthermore, the decision response time was not strongly correlated with seniority for all four conditions $(F_{1,18} = 0.98$, ns). 
Figures \ref{fig:correct-answers} b) and \ref{fig:correct-answers} c) illustrate these results graphically.



\vspace{0.2cm}

\noindent Following the user study, we conducted post-experiment interviews with all participants.
Three users responded that they did not have any trouble identifying correct placements of virtual organs in both the NRM and RM conditions.
Uniformly, all the others stated that the NRM conditions were more intuitive while the RM conditions required more mental workload.

\begin{table*}[tb]
		\centering
		\caption{Comparison of average correct answers and decision times among all 20 participants for all conditions of experiment 1 and 2. The NRM conditions provide a higher overall average of correct answers with slightly smaller decision times compared to the RM conditions. Conditions where organs were flipped correspond to anatomically incorrect organ placements.}
		\vspace{+5mm}
		\renewcommand{\arraystretch}{1.7}
		\centering
		\begin{tabularx}{\textwidth}{@{}l c   cc c   cc}
			\toprule
			
			\textbf{Conditions} & \phantom{a} &
			\multicolumn{2}{c}{~~~~~~~~\textbf{Avg. Correct Answers}~~~~~~~~} & \phantom{a} &
			\multicolumn{2}{c}{~~~~\textbf{~~~~~Avg. Decision Times~~~~~}~~~~}\\
			
			~ & ~ & 
			Mean $\mu$ & SD $\sigma$ & ~ & 
			Mean $\mu$ & SD $\sigma$\\
			
			\cmidrule{1-1} \cmidrule{3-4} \cmidrule{6-7}
			
            \textbf{(STW-NF)}~ See-Through Window, Organs Not Flipped & &
			~~~$94.50 \%$~~~~& $9.46 \%$ & &  
			3.78 $s$ & 2.06 $s$\\
            
            \textbf{(STW-F)}~~~~ See-Through Window, Organs Flipped & &
			~~~$92.00 \%$~~~~& $12.40 \%$ & &  
			3.85 $s$ & 2.07 $s$\\
            
            \\
            
			\textbf{(NRM-NF)}~ Non-Reversing Mirror, Organs Not Flipped & &
			~~~$92.50 \%$~~~~& $10.20 \%$ & &  
			4.50 $s$ & 1.77 $s$\\
			
			\textbf{(NRM-F)}~~~~ Non-Reversing Mirror, Organs Flipped &  &
			$92.00 \%$ & $12.40 \%$ & & 
			~~~~~~5.03 $s$~~~~~~& 2.57 $s$\\
			
			\textbf{(RM-NF)}~~~~ Reversing Mirror, Organs Not Flipped &  &
			$74.00 \%$ & $27.41 \%$ & & 
			5.29 $s$ & 3.39 $s$\\
		
			\textbf{(RM-F)}~~~~~~~ Reversing Mirror, Organs Flipped & &
			$77.00 \%$ & $27.16 \%$ & & 
			5.70 $s$ & 3.39 $s$\\
			
			\bottomrule
			\vspace{+1mm}
				\vspace{-10mm}
		\end{tabularx}
		\label{tab:evaluation_results}
	\end{table*}

%% file: 5_discussion.tex
\section{Discussion \& Future Work}
Our results show that an NRM provides clear benefits over a regular RM design in the area of anatomy
teaching. Participants performed significantly better in identifying the correct placement of virtual anatomical structures when their true mirror
image was shown in form of a NRM, therefore confirming hypothesis 1. 
The average of correct answers for the NRM conditions in experiments 2 were even comparable to the see-through window conditions of experiment 1, which resembled a patient examination and textbook anatomy view medical students are primed to. These results also demonstrate the potential of an AR anatomy teaching system following the NRM paradigm. Medical
students are capable of quickly adapting to such a view, which could be due to the mere-exposure effect \cite{Burgess1971, Mita1977}, a psychological phenomenon by which a person develops a strong preference for a certain stimulus through continuous exposure. Therefore, a RM visualization is not consistent with what they usually observe.
Another interesting usage scenario are student-teacher interactions. Assuming a teacher is explaining anatomy to students by means of his own body in front of a Magic Mirror AR system, the students can develop a much better understanding of the anatomy and link it to their previously acquired knowledge from textbooks with an NRM system.
Regarding hypothesis 2, average decision times among participants were also slightly smaller for the NRM, though we did not observe significant differences.
Throughout the user study, five different virtual organs were traversed for the different conditions. Our statistical analysis did not
reveal significant differences between these organs, showing that overall results were not influenced by one organ performing
better or worse compared to the others.
Hence, we could confirm hypothesis 3. Lastly, we had to reject hypothesis 4, as senior participants 
performed significantly worse for the RM conditions than junior participants.
These results could also be related to the previously discussed mere-exposure effect, as senior medical students have been exposed to the NRM view throughout their studies and multiple patient examinations. 
In future work, we plan to measure the impact of participants' visuo-spatial reasoning skills on both the percentage of correct answers and decision times.
Performance during mental rotation tests such as the one proposed by Shepard and Metzlar could be correlated to how well participants perform in our user study \cite{shepard1988mental}. 
Especially in the field of radiology and orthopedics, spatial ability is important as x-ray images or entire CT and MRI volumes have to be mentally processed and rotated to match the patient.
Participants from such disciplines would be an interesting control group for a follow-up user study.
Furthermore, the development of guidelines for the design of NRM systems for screen-based AR is part of our future research. Now that the potential of NRM designs for screen-based AR application in the field of anatomy learning has been demonstrated, it is intriguing to discuss
the consequences those results have on other fields of AR research. One that immediately comes to mind is interaction. 
In our user study, we did not test for interaction differences as participants made decisions by means of a hand held presenter.
Designing user interfaces for NRM systems should take the reversed interaction into consideration.
Another alluring topic to discuss is whether the results from our user study can be transferred to other disciplines. Our chosen scenario was special in a sense that although RM systems have been applied previously in anatomy learning environments, a mirror view is not the most natural way to present anatomy data.
In a shopping environment for example, people are accustomed to examining clothes in a mirror.
We believe that NRM systems should be considered in scenarios where directions (i.e. left and right) are of crucial importance and for which domain-specific pre-knowledge and the mere-exposure effect play an important role.

%% file: 6_conclusion_future_work.tex
\section{Conclusion}
In this paper, we have explored the potential of non-reversing Magic Mirror systems in the context of anatomy learning and discussed perceptual benefits such a design could privde over traditional reversing Magic Mirror systems.
While the latter present a virtual enantiomorph to the user, a non-reversing Magic Mirror shows the true mirror image of a person standing in front of the system, such that left and right are not reversed.
\vspace{0.2cm}

\noindent We conducted a first preliminary user study comparing both visualizations to each other. Non-reversing mirrors proved to be the more natural choice for the task of identifying the correct placement of virtual organs. Medical students achieved higher percentages of correct answers while simultaneously needing less time for their decision. This coincided with participants qualitative opinions, who found that the non-reversing Mirror visualization is a much better fit for the context of AR anatomy learning. 
In this work we have shown that it is important for application designers to consider the intent of their application and the type of users before deciding on the type of mirror visualization for their system.
Previously acquired domain knowledge, as in the case of anatomy learning, as well as the mere-exposure effect, can make a non-reversing Magic Mirror design the better choice for screen-based AR applications.